\documentclass[a4paper,fleqn,usenatbib]{mnras}

\usepackage{newtxtext,newtxmath}

\usepackage[T1]{fontenc}
\usepackage{ae,aecompl}

\usepackage{graphicx}   
\usepackage{amsmath}    
\usepackage{amssymb}    
\usepackage{longtable}
\usepackage{multirow}
\usepackage{lipsum}
\usepackage{rotating}
\usepackage{bm}
\usepackage{color}
\usepackage[T1]{fontenc}
\usepackage{ae,aecompl}
\usepackage{booktabs,caption}
\usepackage[flushleft]{threeparttable}
\usepackage{subfigure}
\hypersetup{draft}

\title[Statistical properties of SGRs and FRBs]{Statistical properties magnetar bursts and FRB 121102}

\author[Yingjie Cheng, G. Q. Zhang and F. Y. Wang]{
    Yingjie Cheng,$^{1}$
    G. Q. Zhang,$^{1}$
    F. Y. Wang$^{1,2}$\thanks{E-mail: fayinwang@nju.edu.cn}
    \\
    $^{1}$School of Astronomy and Space Science, Nanjing University, Nanjing 210093, China\\
    $^{2}$Key Laboratory of Modern Astronomy and Astrophysics (Nanjing University), Ministry of Education, Nanjing 210093, China
}

\date{Accepted XXX. Received YYY; in original form ZZZ}

\begin{document}
    \label{firstpage}
    \pagerange{\pageref{firstpage}--\pageref{lastpage}}
    \maketitle

\begin{abstract}
In this paper, we present statistics of soft gamma repeater (SGR)
bursts from SGR J1550-5418, SGR 1806-20 and SGR 1900+14 by adding
new bursts from K{\i}rm{\i}z{\i}bayrak et al. (2017) detected with
the Rossi X-ray Timing Explorer (RXTE). We find that the fluence
distributions of magnetar bursts are well described by power-law
functions with indices 1.84, 1.68, and 1.65 for SGR J1550-5418, SGR
1806-20 and SGR 1900+14, respectively. The duration distributions of
magnetar bursts also show power-law forms. Meanwhile, the waiting
time distribution can be described by a non-stationary Poisson
process with an exponentially growing occurrence rate. These
distributive features indicate that magnetar bursts can be regarded
as a self-organizing critical process. We also compare these
distributions with the repeating FRB 121102. The statistical
properties of repeating FRB 121102 are similar with magentar bursts,
combing with the large required magnetic filed ($B\geq 10^{14}$G) of
neutron star for FRB 121102, which indicates that the central engine
of FRB 121102 may be a magnetar.
\end{abstract}


\section{Introduction}
\label{sec:intro}
Soft gamma repeaters (SGRs) are characterized by
short ($\sim 0.1s$) recurrent emission of gamma-rays and X-rays at
irregular intervals (Kouveliotou 1995; Turolla et al. 2015;
Kaspi \& Beloborodov 2017). Different from normal gamma-ray bursts,
the repeated bursts come from the same object and the photons are
less energetic in soft gamma-ray and hard X-ray band. It has long
been accepted that SGRs come from the dissipation of magnetic energy
of an extremely magnetized neutron star (Thompson \& Duncan
1995; Woods, \& Thompson 2006; Mereghetti 2008). The ultra-strong
magnetic fields rotating with the central engine cause great stress
to build up through the crust. When the neutron star crust could no
longer support the stress, it fractures and produces the so-called
`starquakes'. Finally, the ejecting high-energy particles are
captured by the near-surface magnetic field, causing emissions of
soft gamma rays and X-rays. Thompson et al. (2002) proposed that the
magnetospheres of neutron stars are globally twisted. Lyutikov
(2003) suggested that bursts may be produced by magnetic
reconnection in the magnetosphere of magnetars.

Recent studies have found that the distribution of SGR burst energy
can be well fitted with a power-law function
\citep{Gogus1999,2000ApJ...532L.121G,2001ApJ...558..228G,
2007PASJ...59..653N,2008ApJ...685.1114I,2011ApJ...740L..16L,2012ApJ...749..122V,2012ApJ...755....1P}.
This power-law distribution of energy release has been seen in many
natural systems featuring nonlinear energy dissipation and is
commonly regarded as a symbol of self-organized critical (SOC)
systems \citep{1987PhRvL..59..381B,2011soca.book.....A}. First
introduced by Bak, Tang, and Wiesenfeld in 1987
\citep{1987PhRvL..59..381B}, the concept of SOC has been applied to
a wide variety of astrophysics (Lu \& Hamilton 1991; Gogus et al.
1999; Aschwanden 2011; Wang \& Dai 2013; Wang et al. 2015; Li et al.
2015; Harko et al. 2015; Yi et al. 2016; Wang, Wang \& Dai 2017;
Wang \& Yu 2017; Yan et al. 2018; Zhang et al. 2019). SOC refers to
a critical state with instability threshold in a nonlinear energy
dissipation system. The great success of \citet{1991ApJ...380L..89L}
in explaining solar flares by an SOC system made SOC a widely
popular concept. Earthquakes, sharing many statistical properties
with SGR bursts, have been described by a SOC system
\citep{1989JGR....9415635B,1992PhRvL..68.1244O}. Inspired by the
earthquakes models, a SOC interpretation of SGR bursts was derived
\citep{1996Natur.382..518C}.

Many researches on the statistical characteristics of SGRs have been
carried out, most of which focus on the distribution of burst energy
\citep{1996Natur.382..518C,Gogus1999,2000ApJ...532L.121G,2012ApJ...755....1P}.
The energy or total counts of SGR bursts are proved to have
power-law-like size distributions. The frequency
distributions of duration and waiting time are also important
features of SOC systems (Aschwanden 2011). However, they have not
been investigated for SGRs. Recently, K{\i}rm{\i}z{\i}bayrak et al.
(2017) presented broadband (2-250 keV) time-averaged spectral
analysis of a large sample of bursts from SGR J1550-5418, SGR
1900+14, and SGR 1806-20 detected with the Rossi X-ray Timing
Explorer (RXTE). Our analysis is based on these public data.

Fast radio bursts (FRBs) are milliseconds mysterious radio
transients with anomalously high dispersion measure
\citep{2007Sci...318..777L,2013Sci...341...53T,Petroff19}. FRBs are
believed to occur at cosmological distances, which is supported by
the direct localization of FRB 121102 (Chatterjee et al. 2017), FRB
180924 (Bannister et al. 2019), and FRB 190523 (Ravi et al. 2019).
Ten FRBs have been found to be repeating (Spitler et al. 2016;
CHIME/FRB Collaboration 2019a,b). The central engine of FRBs is
still a mystery. Many theoretical models invoke magnetars (Platts et
al. 2018; Cordes \& Chatterjee 2019, and references therein). FRB
121102 has an extremely high Faraday rotation measure, indicating
strongly magneto-ionic surrounding medium (Michilli et al. 2018),
combing a coincidental continuum radio source (Chatterjee et al.
2017; Marcote et al. 2017), has inspired a model involving a young
magnetar in an expanding supernova remnant (Metzger et al. 2017; Cao
et al. 2017; Metzger et al. 2019).

In this paper, we give statistical analysis of over 1,500 bursts
from SGR J1550-5418, SGR 1806-20 and SGR 1900+14 observed by the
RXTE between 1996 and 2011. The cumulative distributions of total
counts, burst duration and waiting time are shown in section
\ref{sec:data}. The distributions of total counts and burst duration
show a threshold power-law form, while the waiting time distribution
can be well described by a non-stationary Poisson process with an
exponentially growing occurrence rate. The best-fitting results for
each distribution can be found in section \ref{sec:result}. We
compare the distributions between magnetar bursts and FRB 121102 in
section 4. The conclusions are given in section \ref{sec:discuss}.

\section{Data and Methods}
\label{sec:data}

According to the updated catalog offered by the McGill Pulsar Group,
a total of 15 SGRs (11 confirmed, 4 candidates) have been found
\citep{2014ApJS..212....6O}. The most well-studied sources include
SGR J1550$-$5418, SGR 1806$-$20 and SGR 1900$+$14.
\citet{1995Ap&SS.231...49K} have built an on-line database of
magnetar bursts observed by the Rossi X-ray Timing Explorer (RXTE)
which can be found at \url{http://magnetars.sabanciuniv.edu.} In
this paper, we consider 1,535 bursts from SGR J1550-5418, SGR
1806-20 and SGR 1900+14 detected in 15 years. A short description of
the database entries is given in Table \ref{tab:obs}. We focus on
the distributions of fluence, duration and waiting time of magnetar
bursts. The value of fluence and duration can be directly obtained
from the on-line database. Power-law distribution of duration is
also an important prediction of SOC theory
\citep{1987PhRvL..59..381B,2011soca.book.....A}. However, this
prediction has not been explored for magnetar bursts. The frequency
distribution of duration is studied is this paper. We have measured
the waiting times between successive bursts through
$\delta_t=t_{start,i+1}-t_{start,i}$, where $t_{start,i+1}$ and
$t_{start,i}$ are the start times for the $i+1$th and $i$th bursts,
respectively. However, in order to avoid the effects of Earth
occultation and data gaps, only the continuous observation data is
used.

\begin{table}
 \centering
 \caption{\label{tab:obs}}
 \begin{tabular}{ccc}
  \toprule
  Source Name & Observation Period & Burst Number\\
  \midrule
  SGR J1550-5418 & Oct 2000-Apr 2010 & 179\\
  SGR 1806-20 & Nov 1996-Jun 2011 & 924\\
  SGR 1900+14 & Jun 1998-Apr 2006 & 432\\
  \bottomrule
 \end{tabular}
\begin{tablenotes}
    \item List of the database entries.
\end{tablenotes}
\end{table}

Generally, a nonlinear process can be simply expressed by an
exponential growth phase with saturation after a random time
interval, which has been applied to many different scientific areas
\citep{1922Natur.109..177W,1978ApJ...222.1104R,2005ConPh..46..323N,2012CNSNS..17.3558P}.
The size distribution \textit{N(x)} of exponentially growing
avalanches can be written as
\begin{equation}
\label{equ:1}
N(x)dx = n_{0}(x_{0}+x)^{-\alpha_x}dx.
\end{equation}
For $x_{1} \leqslant x \leqslant x_{2}$, the normalization constant
$n_{0}$ is expressed as
\begin{equation}
\label{equ:2}
n_{0} = n_{ev}(1-\alpha_x)[(x_{0}+x_{2})^{1-\alpha_x}-(x_{0}+x_{1})^{1-\alpha_x}]^{-1}.
\end{equation}
Since a constant $x_{0}$ has been added to the ideal power-law
distribution function, such a distribution can be called as
thresholded power-law size distribution, or differential occurrence
frequency distribution
\citep{1997WRR....33.1567B,2015ApJ...814...19A}. Hereafter, we
simply refer it as the differential distribution.

When the data sample is too small, a cumulative distribution
function is more often used in fitting. The cumulative size
distribution is defined as the integral number of events above a
given value \textit{x}. If $x_{1}$ and $x_{2}$ represent the minimum
and maximum value of the size distribution, and $n_{ev}$ refers to
the total number of events, the thresholded cumulative size
distribution can be expressed as
\begin{equation}
\label{equ:3}
  \begin{split}
  & N_{cum}(>x)dx = \int_x^{x_{2}} n_{0}(x_{0}+x)^{-\alpha_x}dx \\
  & = 1+(n_{ev}-1)(\frac{(x_{0}+x_{2})^{1-\alpha_x}-(x_{0}+x)^{1-\alpha_x}}{(x_{0}+x_{2})^{1-\alpha_x}-(x_{0}+x_{1})^{1-\alpha_x}}).
  \end{split}
\end{equation}
It can be seen that $x_{0}$ and $\alpha_{x}$ are the only two free variables in the cumulative distribution.

The parameter $x_{0}$ can largely improve the fitting result at the
lower bound of a power-law-like distribution, and has its own
physical meanings. It could be attributed to the instability
threshold of the system, as well as the sub-sampling below the
detection threshold \citep{2015ApJ...814...19A}. These two causes
are physically different, but can be mathematically treated in the
same way.

\begin{figure*}
\centering
\subfigure{
\label{fig:subfig:d}
\includegraphics[width=3in]{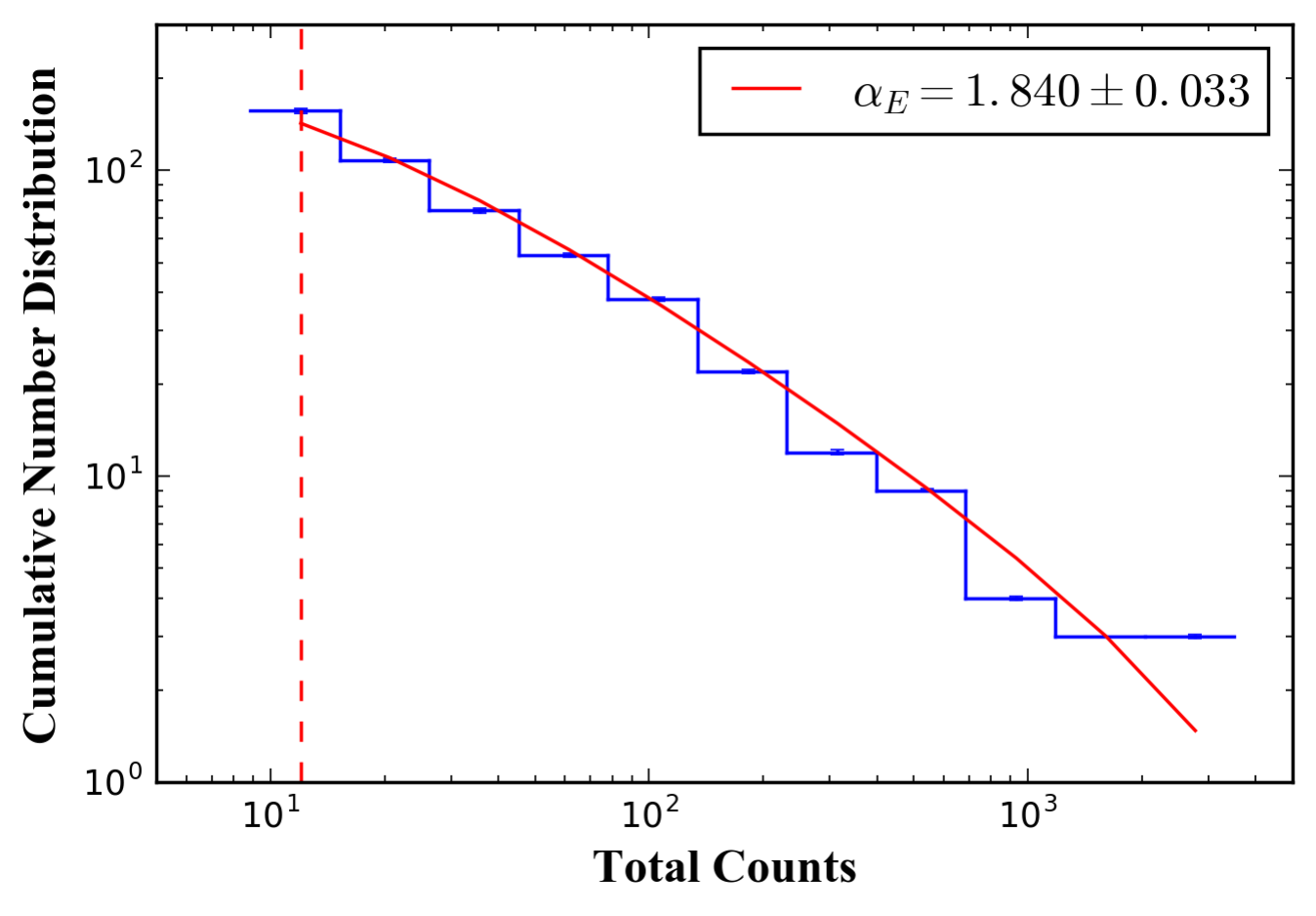}}
\subfigure{
\label{fig:subfig:e}
\includegraphics[width=3in]{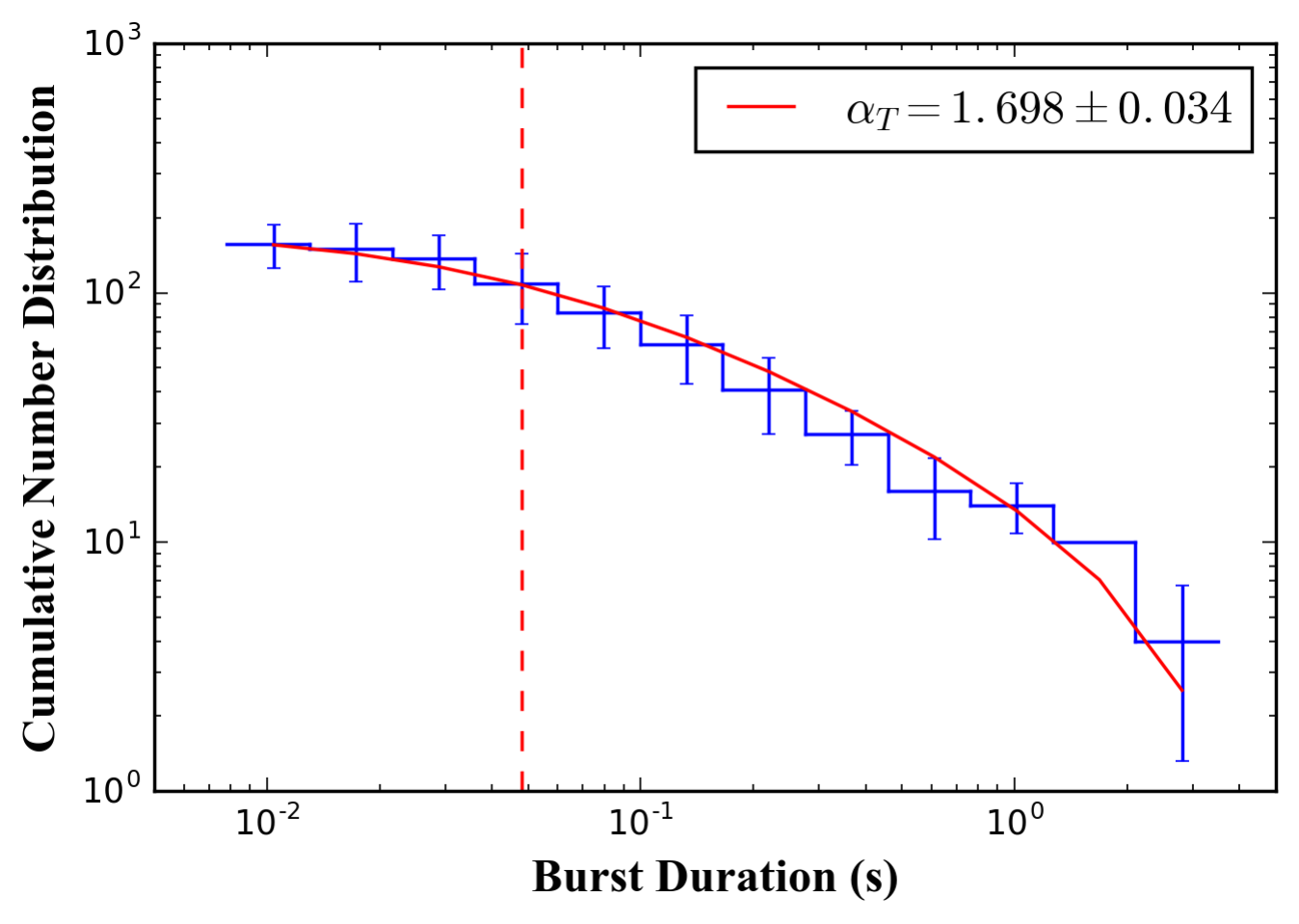}}
\subfigure{
\label{fig:subfig:f}
\includegraphics[width=3in]{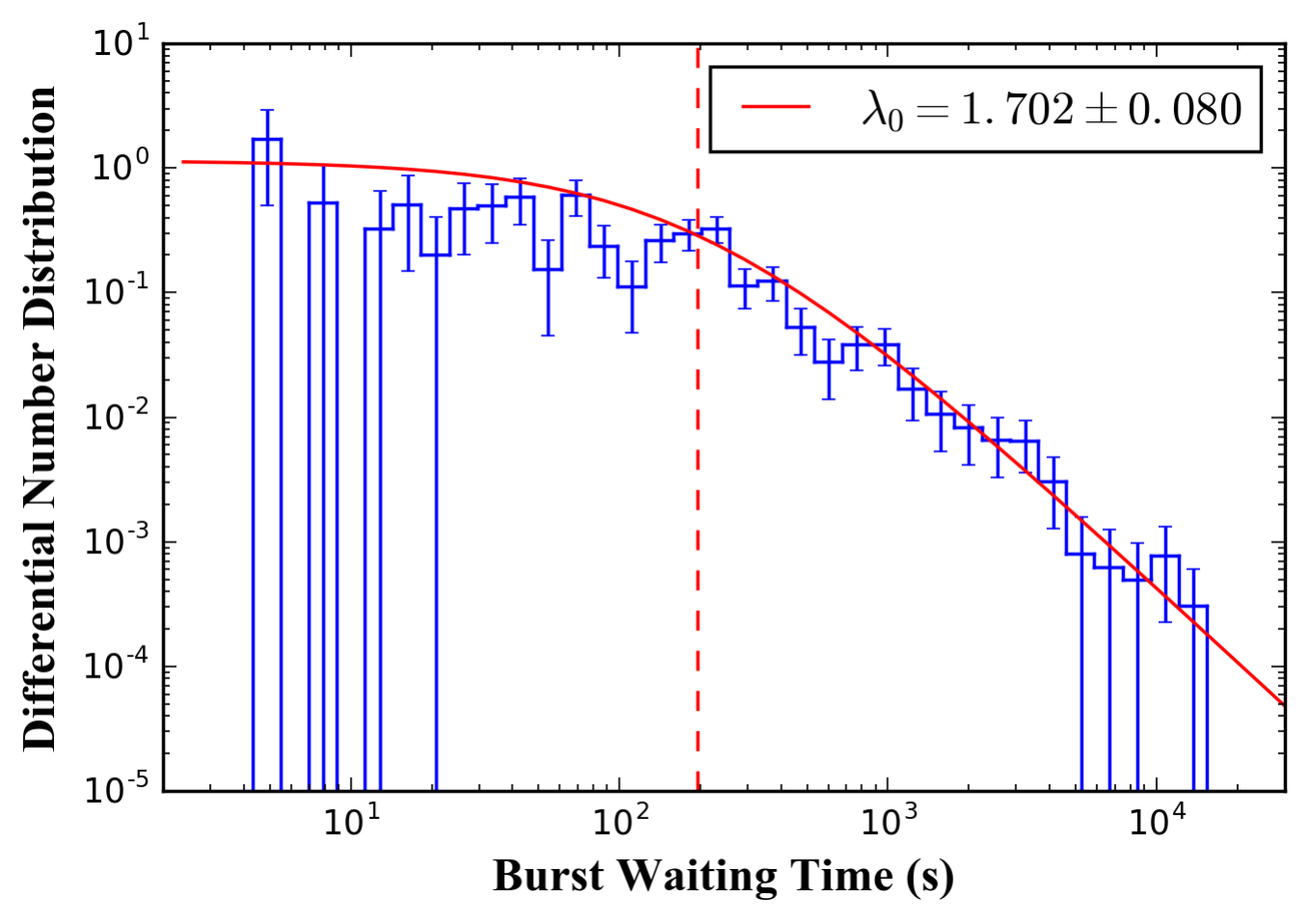}}
\centering
\caption{Distributions of total counts, durations and waiting times and best-fitting results (reds lines) for SGR J1550-5418.}
\label{fig:1}

\subfigure{
\label{fig:subfig:d}
\includegraphics[width=3in]{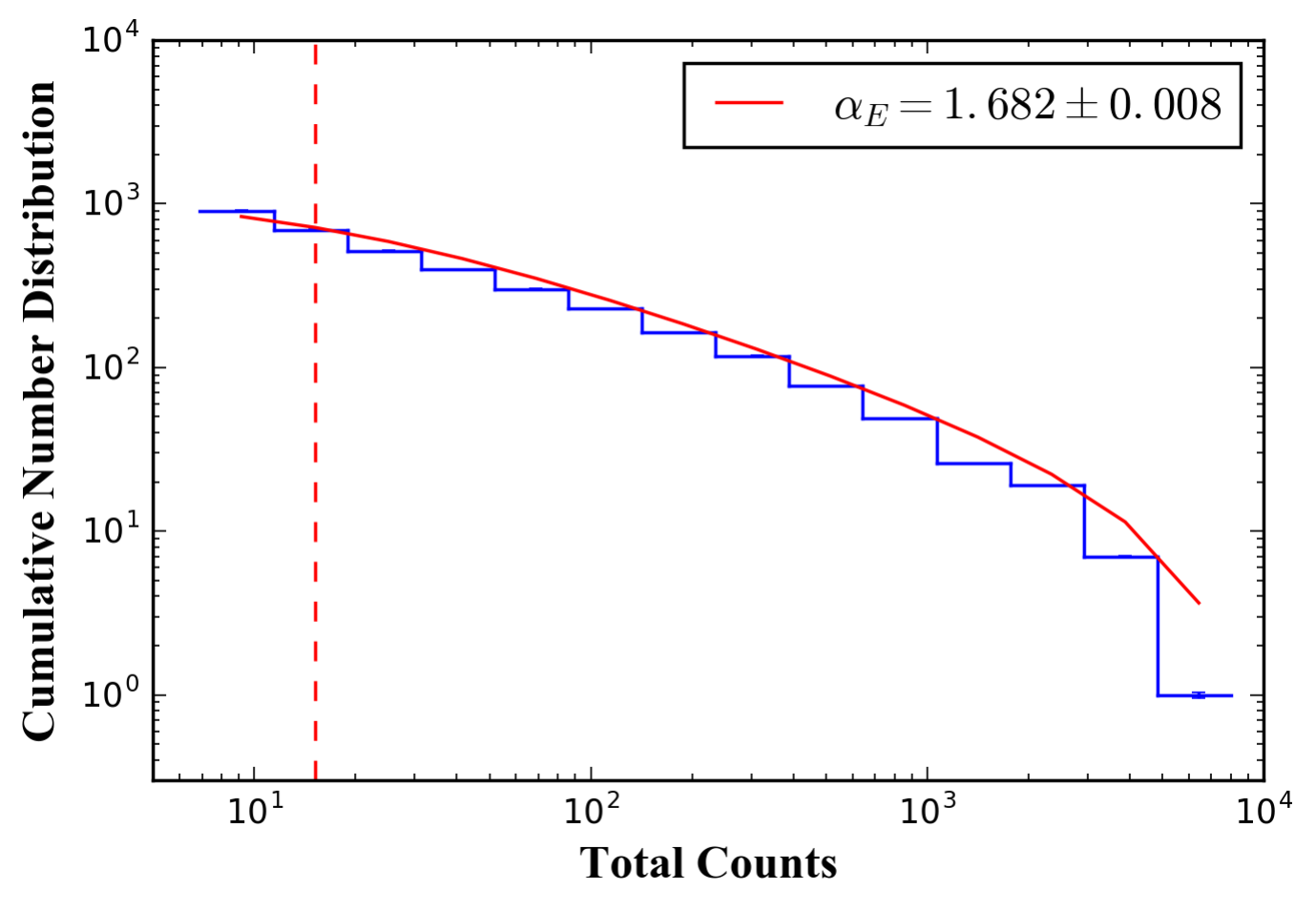}}
\subfigure{
\label{fig:subfig:e}
\includegraphics[width=3in]{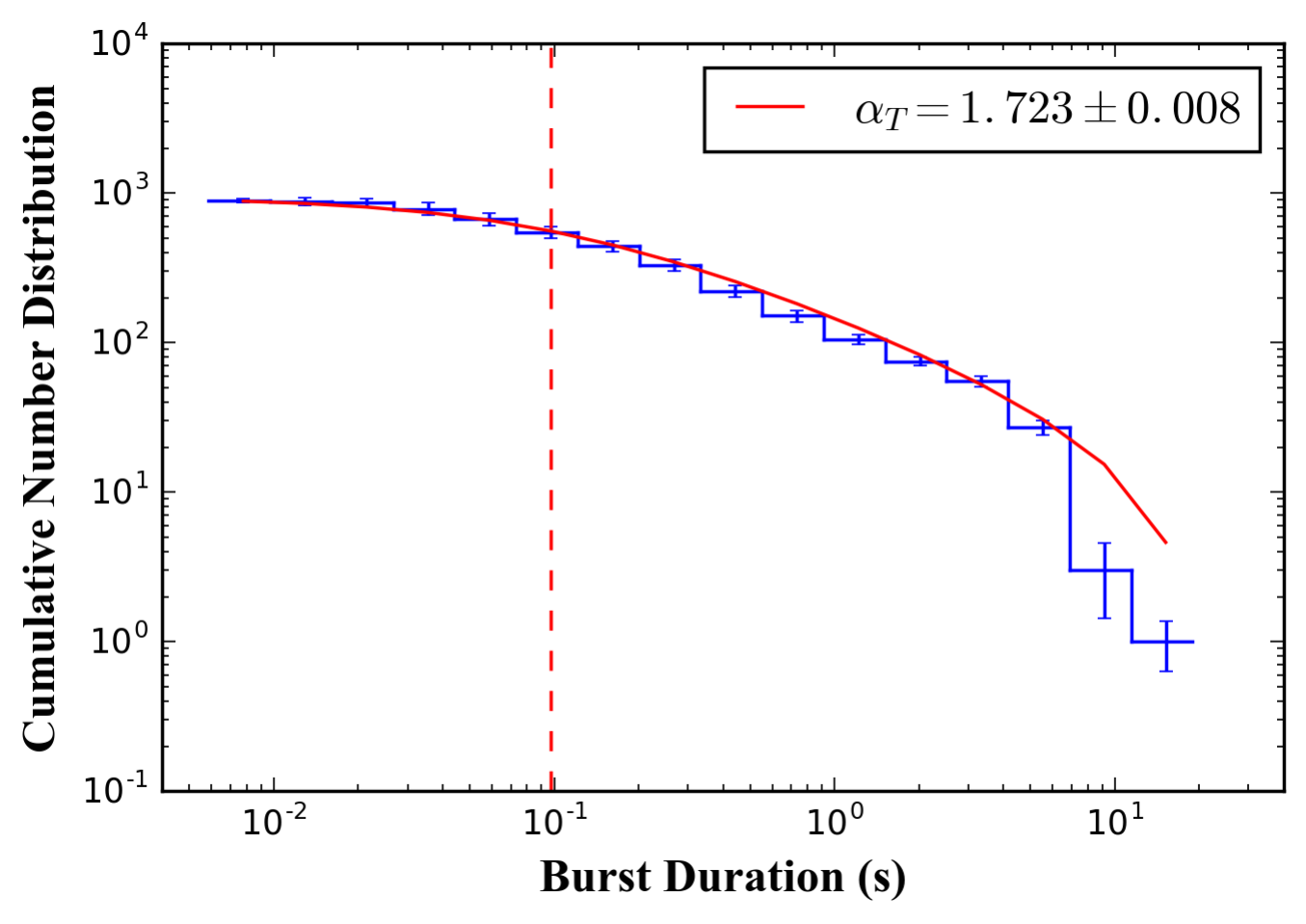}}
\subfigure{
\label{fig:subfig:f}
\includegraphics[width=3in]{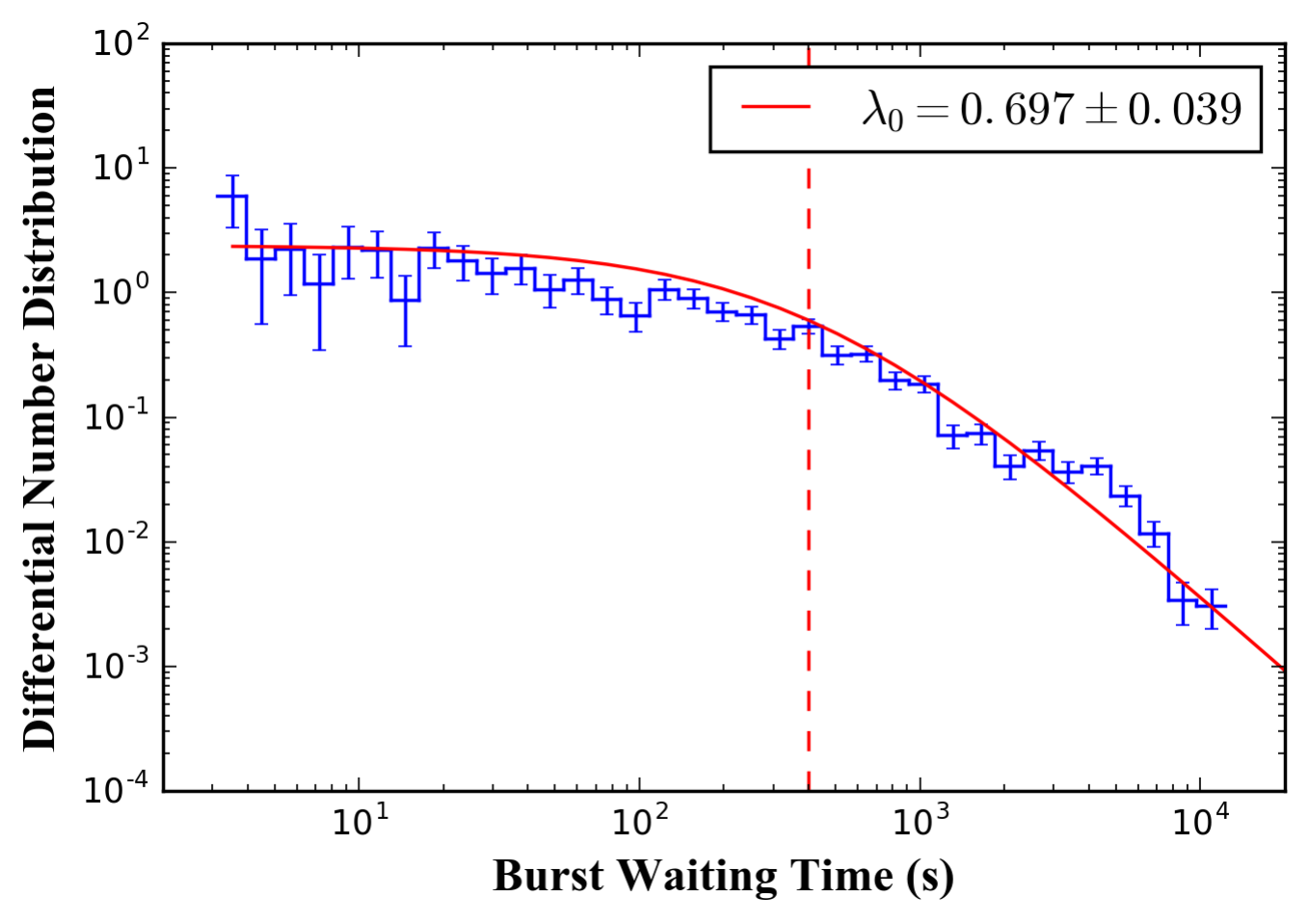}}
\centering
\caption{Distributions of total counts, durations and waiting times and best-fitting results (reds lines) for SGR 1806-20.}
\label{fig:2}
\end{figure*}

For each source, we derived both differential and cumulative
distributions of total counts, burst duration and burst waiting
time. The dataset of each distribution is a series of event sizes
with a total number of $n_{ev}$. To begin with, the data is
uniformly binned on a logarithmic scale between the minimum and
maximum sampled data ($x_{1}$ and $x_{2}$). Empirically, for
differential distributions, we set the number of bins to be
$\lg(x_{2}/x_{1})\times10$, while for cumulative distributions, the
number is $\lg(x_{2}/x_{1})\times5$ \citep{2015ApJ...814...19A}. The
threshold $x_{0}$ is determined from the bin with the maximum number
of events, ensuring that an ideal power-law fit can be applied to
the data range $[x_{0},x_{2}]$. The expected uncertainty of the
differential distribution can be written as
\begin{equation}
\label{equ:4} \sigma_{diff,i} = \sqrt{N_{bin,i}}/\Delta x_{i}.
\end{equation}
While for the cumulative distribution, the uncertainty is
\begin{equation}
\label{equ:5}
\sigma_{cum,i} = \sqrt{N_{bin,i}/\Delta x_{i}}.
\end{equation}
Each data bin as well as the uncertainties are demonstrated in
Figures \ref{fig:1},\ref{fig:2} and \ref{fig:3}). The threshold
$x_{0}$ is marked with a red dashed line.

For the distributions of total counts and burst duration, the
dataset can be well described by Equation (\ref{equ:1}). Since the
total number of events $n_{ev}$ and the constant $x_{0}$ have
already been identified, the normalization constant $n_{0}$ can be
expressed as a function of $\alpha_x$. Thus, the power-law slope
$\alpha_x$ is the only free variable to optimize, which can be
obtained by Bayesian statistical analysis.

For the distributions of burst waiting time, the data reflects a
superposition of multiple exponential distributions with different
time scales, which could be attributed to a non-stationary Poisson
process \citep{1998ApJ...509..448W,2011soca.book.....A}. Such
waiting time distributions indicate random processes with
time-dependent event rates, and can also be characterized with
Bayesian statistics \citep{1998ApJ...509..448W,2002SoPh..211..255W}.
The probability function can be approximately expressed as a
subdivision into discrete time intervals, within which the
occurrence rate is constant (Equation \ref{equ:6}). Thus, the
non-stationary process is regarded as a superposition of several
stationary processes with occurrence rates $\lambda_{1},
\lambda_{2}, \dots, \lambda_{n}$. The waiting time distribution is
\begin{equation}
\label{equ:6}
P(\Delta t)=\left\{\begin{array}{ll}{\lambda_{1} e^{-\lambda_{1} \Delta t}} & {\text { for } t_{1} \leq t \leq t_{2}} \\ {\lambda_{2} e^{-\lambda_{2} \Delta t}} & {\text { for } t_{2} \leq t \leq t_{3}} \\ {\ldots \ldots \ldots} & {\text { for } t_{n} \leq t \leq t_{n+1}} \\ {\lambda_{n} e^{-\lambda_{n} \Delta t}} & {\text { for } t_{n} \leq t \leq t_{n+1}}\end{array}\right.
\end{equation}

The dataset shows power-law-like distributions with slopes close to
2 for large waiting times. We suppose that the variability of burst
rate shows spikes like $\delta-$functions (Aschwanden \& McTiernan
2010). The occurrence rate $\lambda$ is exponentially growing
(equation \ref{equ:7}) and fulfills the normalization requirement
$\int_{0}^{\infty} \lambda f(\lambda) d \lambda=\lambda_{0}$
\begin{equation}
\label{equ:7} f(\lambda)=\lambda^{-1} \exp
\left(-\frac{\lambda}{\lambda_{0}}\right).
\end{equation}
The waiting time distribution in a given time interval can be written as
\begin{equation}
\label{eq:Pdt} P(\Delta t)=\frac{\lambda_{0}}{\left(1+\lambda_{0}
\Delta t\right)^{2}},
\end{equation}
where the mean burst rate $\lambda_{0}$ is the only variable to optimize.

\vspace{5mm}
\section{Fitting Results}\label{sec:result}
We use the open source probabilistic programming framework PyMC3 to
perform the fitting. PyMC3 uses Theano to compute gradients via
automatic differentiation and allows model specification directly in
Python code \citep{2016ascl.soft10016S}. For the distributions of
total counts and burst duration, the power-law slope $\alpha_x$ is
created as a stochastic random variable with normal prior
distribution and the step method is run for 5000 iterations. The
mean value and standard deviation of the collected samples in the
returned trace object are regarded as the best-fit slope and its
uncertainty for each distribution.

For cumulative distributions, the best-fit power-law slopes of total
counts are $1.840\pm0.033$, $1.682\pm0.008$, and $1.654\pm0.014$ for
SGR J1550-5418, SGR 1806-20 and SGR 1900+14, respectively. These
value are consistent with those derived by Gogus et al. (1999, 2000)
and Prieskorn \& Kaaret (2012). SOC theory not only predicts
the power-law distribution of energy, but also predicts power-law
distribution of duration \citep{2011soca.book.....A}. For duration
distributions, the power-law slopes are $1.698\pm0.034$,
$1.723\pm0.008$, and $1.821\pm0.016$ for SGR J1550-5418, SGR 1806-20
and SGR 1900+14, respectively.

\begin{table*}
\renewcommand\tabcolsep{12pt}
\label{tab:2}
  \begin{tabular}{ccc}
  \toprule
  \hline
  \multicolumn{3}{c}{SGR J1550-5418} \\
  \hline
  & Differential Distribution & Cumulative Distribution \\
Total counts ($\alpha_E$) & $1.667\pm0.078$ & $1.840\pm0.033$ \\
Burst Duration ($\alpha_T$) & $1.503\pm0.008$ & $1.698\pm0.034$ \\
Waiting Time ($\lambda_0$) & $1.702\pm0.080~s^{-1}$ & $/$ \\
  \hline
  \multicolumn{3}{c}{SGR 1806-20} \\
  \hline
  & Differential Distribution & Cumulative Distribution \\
Total counts ($\alpha_E$) & $1.741\pm0.067$ & $1.682\pm0.008$ \\
Burst Duration ($\alpha_T$) & $1.430\pm0.005$ & $1.723\pm0.008$ \\
Waiting Time ($\lambda_0$) & $0.697\pm0.039~s^{-1}$ & $/$ \\
  \hline
  \multicolumn{3}{c}{SGR 1900+14} \\
  \hline
  & Differential Distribution & Cumulative Distribution \\
Total counts ($\alpha_E$) & $1.672\pm0.048$ & $1.654\pm0.014$ \\
Burst Duration ($\alpha_T$) & $1.505\pm0.006$ & $1.821\pm0.016$ \\
Waiting Time ($\lambda_0$) & $1.311\pm0.049~s^{-1}$ & $/$ \\
  \hline
  \multicolumn{3}{c}{FRB 121102} \\
  \hline
  & Differential Distribution & Cumulative Distribution \\
  Energy ($\alpha_E$) & - & $1.63 \pm 0.06$ \\
  Duration ($\alpha_T$) & - & $1.57 \pm 0.13$ \\
  Waiting Time ($\lambda_0$) &  $1.23^{+0.80}_{-0.38} \times 10^{-5}~\rm ms^{-1}$ & - \\
  \hline
  \bottomrule
  \end{tabular}
\centering \caption{The best-fit parameters of cumulative
distributions and differential distributions for SGR J1550-5418, SGR
1806-20, SGR 1900+14 and FRB 121102.}
\end{table*}

The waiting time distribution of these three datasets can be
described by a Poisson processes with mean burst rate $\lambda$.
Similarly, the best-fit $\lambda$ and its uncertainty for each
waiting time distribution is obtained by PyMC3. Because observations
cover a long time, the burst rate varies with times, which can be
treated as a non-stationary Poisson process.

\begin{figure*}
\centering
\subfigure{
\label{fig:subfig:d}
\includegraphics[width=3in]{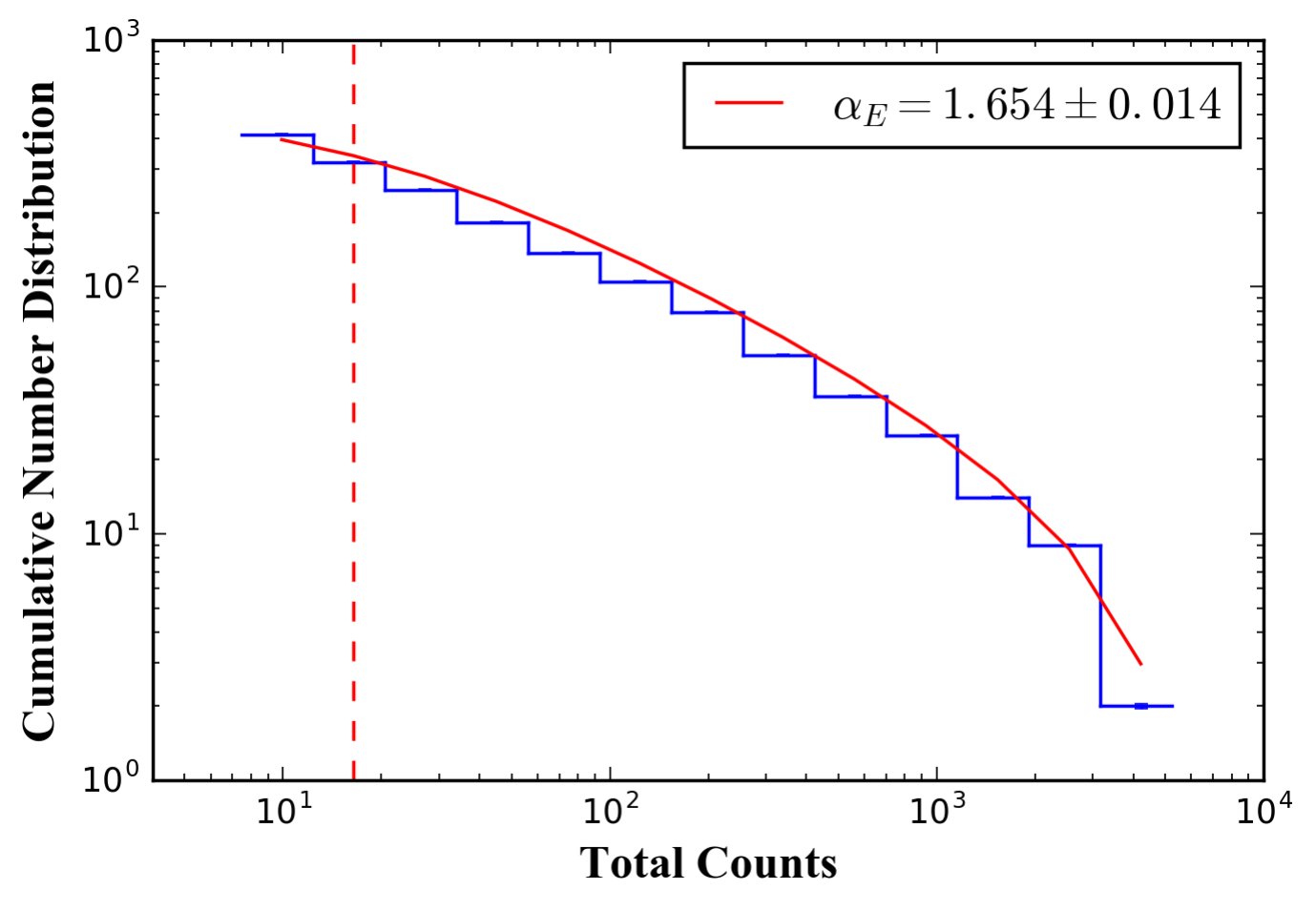}}
\subfigure{
\label{fig:subfig:e}
\includegraphics[width=3in]{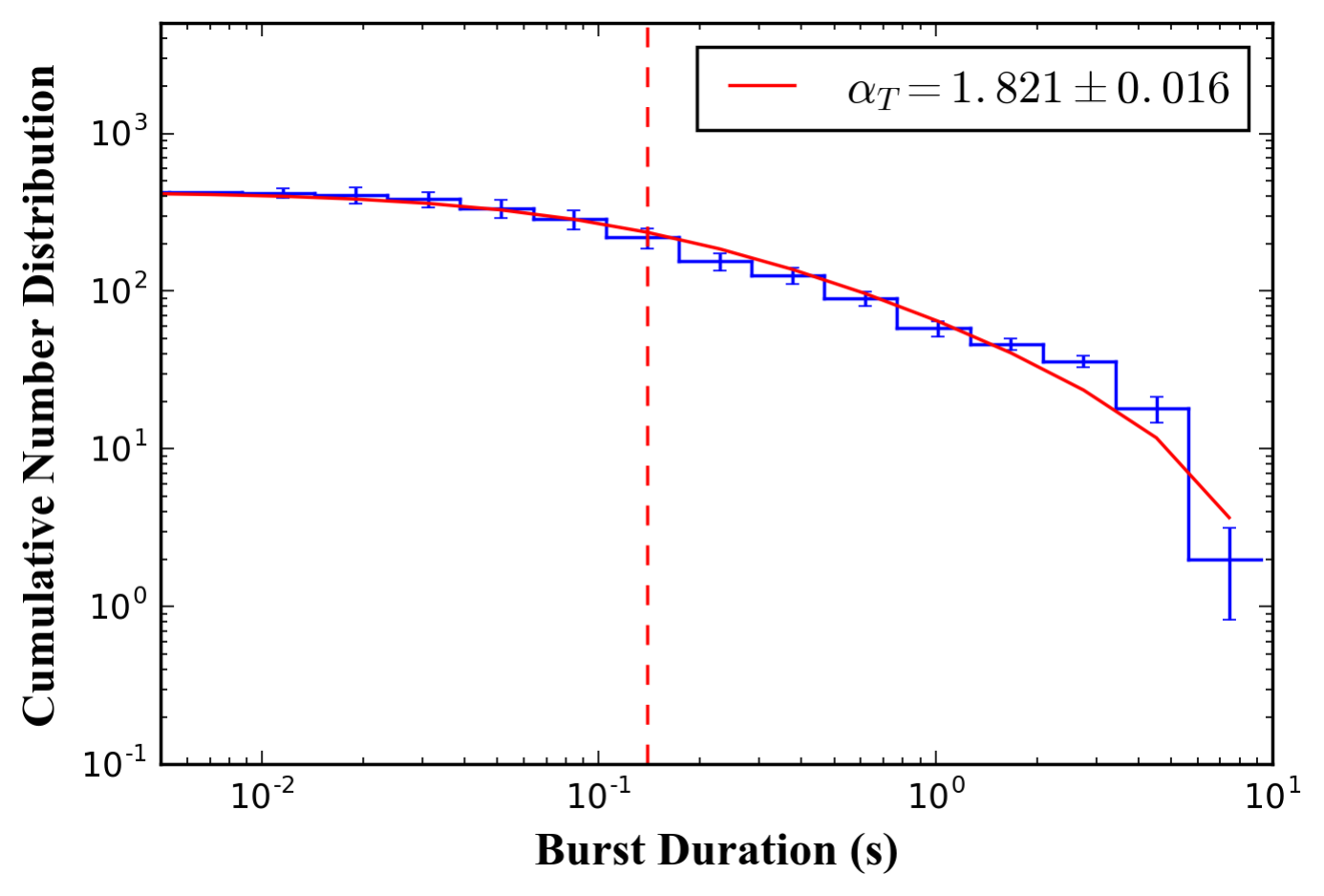}}
\subfigure{
\label{fig:subfig:f}
\includegraphics[width=3in]{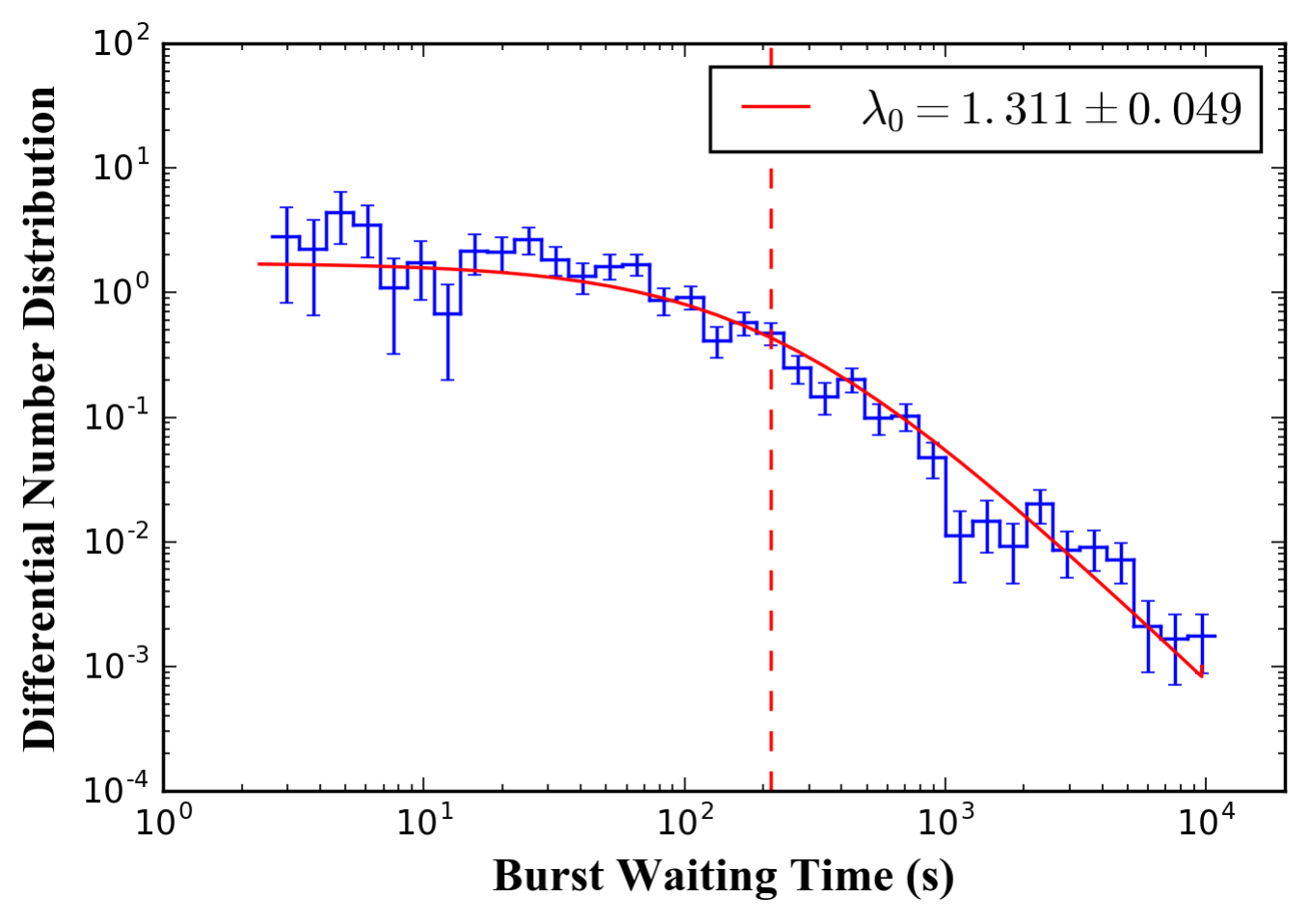}}
\centering
\caption{Distributions of total counts, durations and waiting times and best-fitting results (reds lines) for SGR 1900+14.}
\label{fig:3}

\subfigure{
\label{fig:subfig:d}
\hspace{2mm}\includegraphics[width=3in]{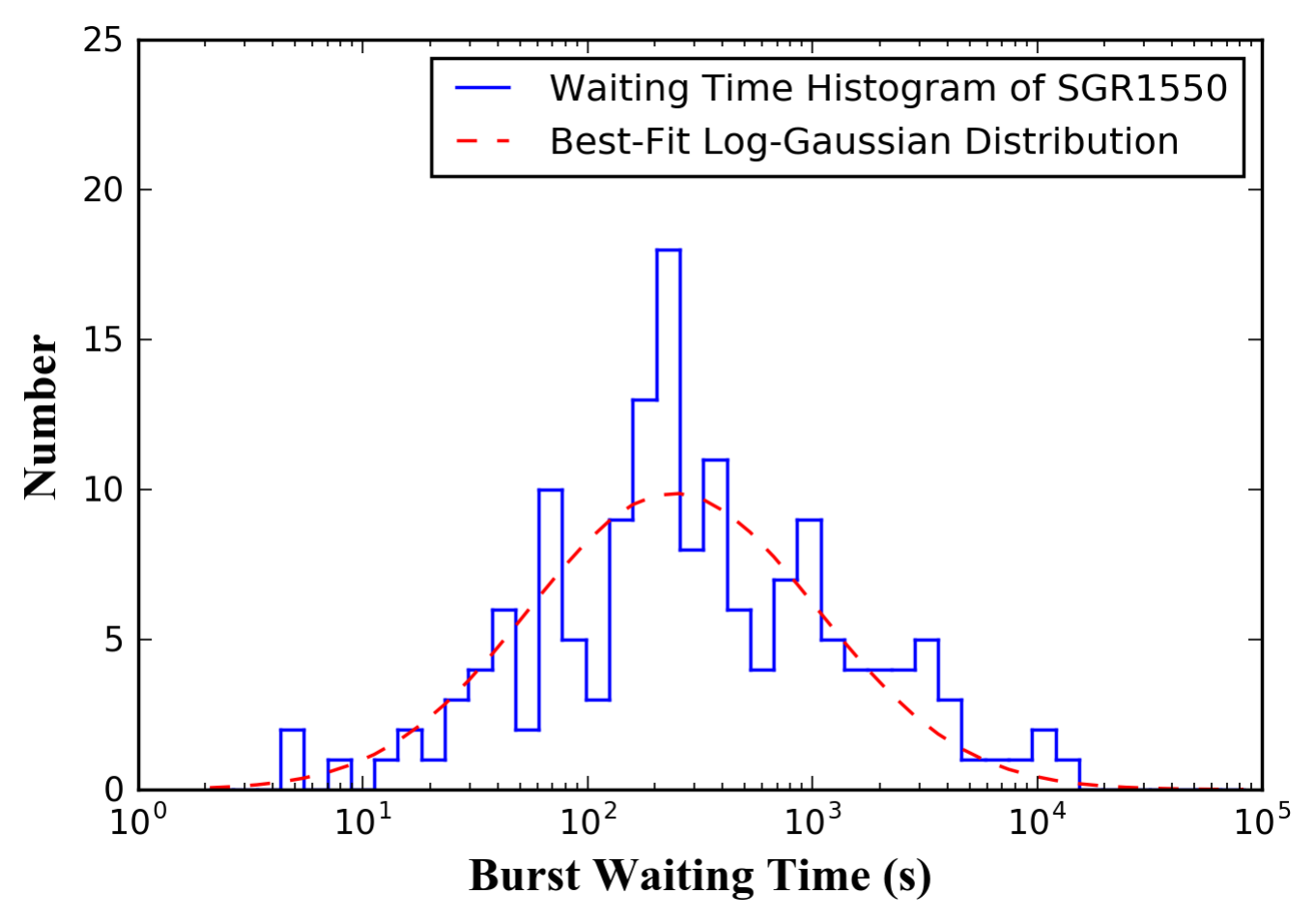}}
\subfigure{
\label{fig:subfig:e}
\includegraphics[width=3in]{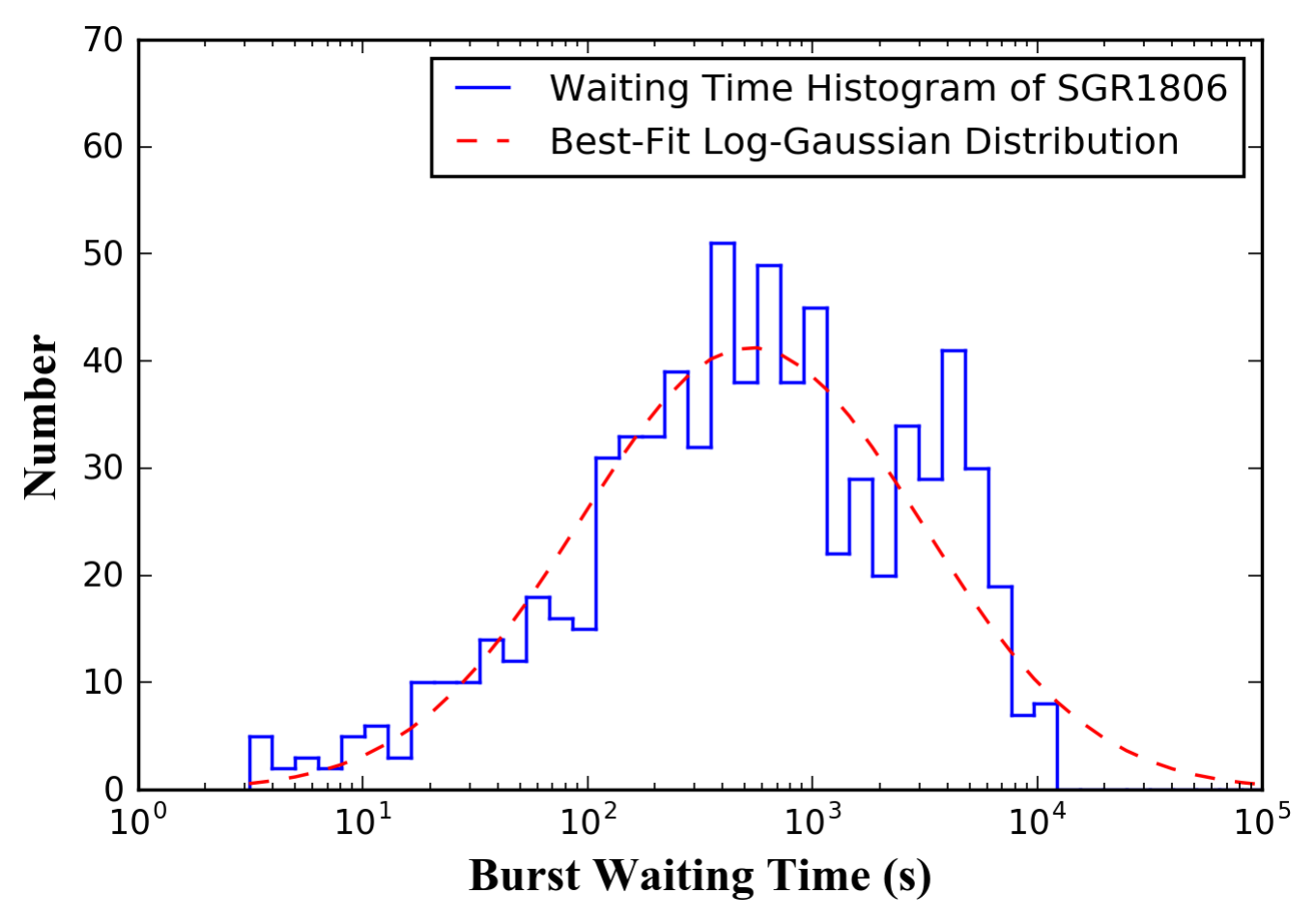}}
\subfigure{
\label{fig:subfig:f}
\hspace{3mm}\includegraphics[width=3in]{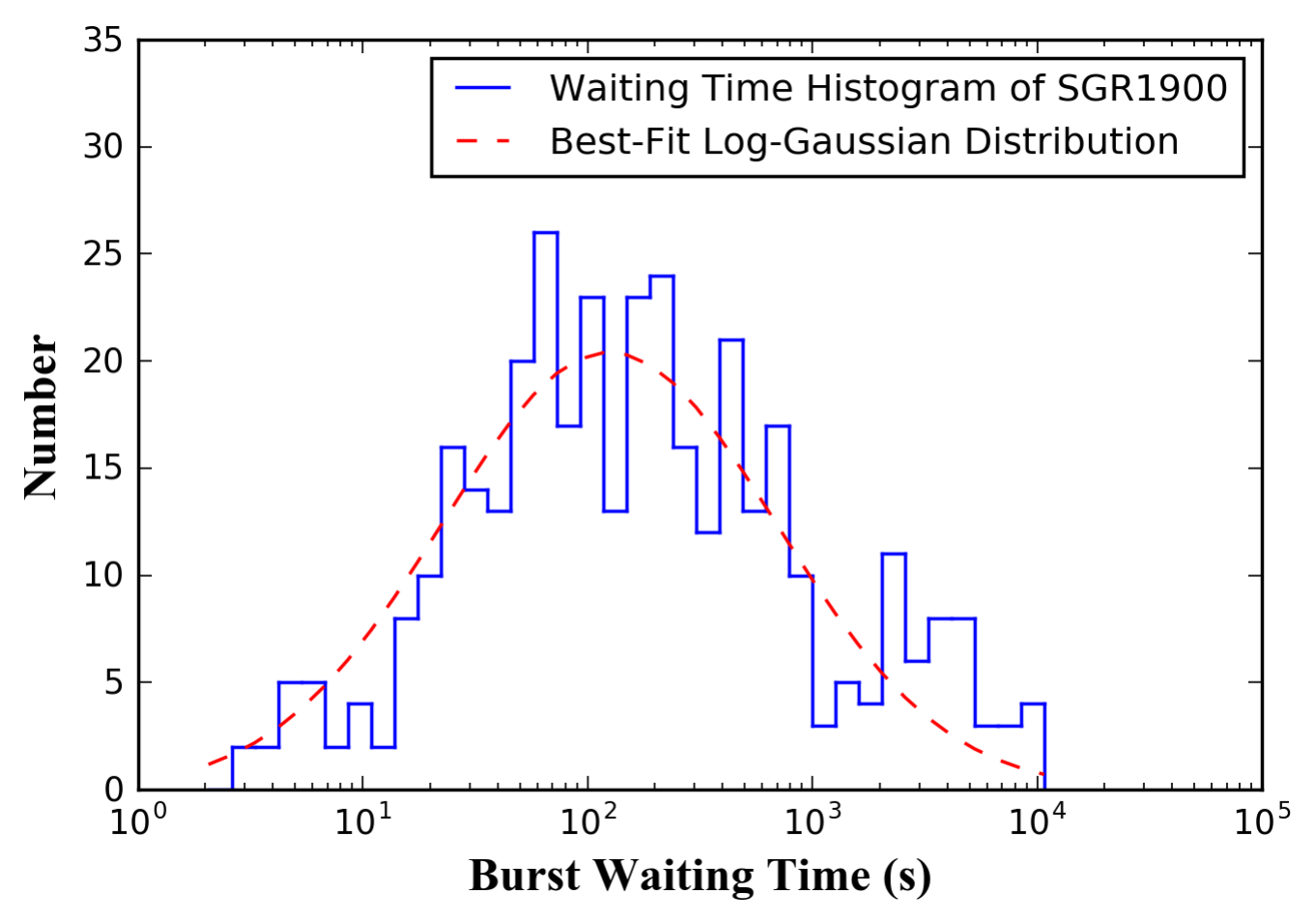}}
\centering
\caption{Frequency distribution histograms of waiting time for SGRs. The best-fit log-Gaussian distribution curves are marked as red dashed lines.}
\label{fig:4}
\end{figure*}

A compilation of the best-fitting parameters for all the
distributions is listed in Table 2. Also, the fits of cumulative
size distributions are shown in Figures \ref{fig:1}, \ref{fig:2} and
\ref{fig:3} for SGR J1550-5418, SGR 1806-20 and SGR 1900+14,
respectively.

Figure \ref{fig:4} shows the histograms of the burst waiting times.
We have fit the waiting time distributions to a log-Gaussian
function. The peaks are 208 s (with $\sigma \sim 4.27$), 478 s
($\sigma \sim 6.31$) and 115 s (with $\sigma \sim 5.75$) for SGR
J1550-5418, SGR 1806-20 and SGR 1900+14, respectively.
\cite{Gogus1999} found that the waiting times of SGR 1900+14
can be fitted with log-Gaussian distribution, but with a low peak
about $49$s. For waiting times of SGR 1806-20, \cite{2000ApJ...532L.121G}
derived a log-Gaussian distribution with a peak at 103 s from
observation and numerical simulations. Considering the errors of
fitting, our results for SGR 1806-20 and SGR 1900+14 are larger than
those of \cite{Gogus1999} and \cite{2000ApJ...532L.121G}. The main reason is
that the difference of burst identification methods. In
\cite{Gogus1999} and \cite{2000ApJ...532L.121G}, the bursts are selected using
a phase-folding technique \citep{Woods1999}. The bursts used in this
paper are analyzed using Bayesian blocks algorithm provided in
\cite{Scargle2013} and \cite{Lin2013}. For SGR 1900+14, 837 bursts
are derived by \cite{Gogus1999}, compared to 432 bursts with
Bayesian blocks algorithm in a long observation period
\citep{kirm2017}. Therefore, the peak of waiting time distributions
is larger for the bursts in \cite{kirm2017}. Interestingly,
\cite{Hurley1994} found that the waiting times of SGR 1806-20 can be
fitted with log-Gaussian distribution with a peak about
$1.63\times10^4$ s. This result may be affected by data gap.
Overall, our result is between the value of \cite{Gogus1999} and
\cite{Hurley1994}.

\section{Comparison with fast radio bursts}
\label{sec:frb}

In this section, we compare the distributions between SGRs and fast
radio bursts (FRBs). By now, dozens of FRBs have been
discovered, which are listed in FRB Catalogue (http://frbcat.org/)
\citep{Petroff16}. Eleven FRBs are repeating
\citep{2016Natur.531..202S,Amiri2019,Andersen2019}. There are some
phenomenological similarities between SGRs and FRBs, including
repeatability, timescales and the duty factor of pulses
\citep{2014ApJ...797...70K,2016ApJ...826..226K,2017JCAP...03..023W}.

The high brightness temperatures $\geq 10^{37}$ K of FRBs
require a coherent emission process (Katz 2016; Lyutikov 2019). The
two most commonly mechanisms are coherent curvature radiation
produced near the surface of the neutron star (Lu \& Kumar 2018;
Yang \& Zhang 2018) and the synchrotron maser process (Lyubarsky
2014; Metzger et al. 2019). For the synchrotron maser
process, the emission is from an ultra-relativistic shock moving
towards observer, which propagates into medium of moderately high
magnetization, $\sigma>10^3$ (Lyubarsky 2014; Beloborodov 2017). In
the magnetar scenario, these shocks result from the sudden release
of energy during the earliest stages of a flare. Some theoretical
models for FRBs basing on magnetars have been proposed (Popov \&
Postnov 2013; Lyubarsky 2014; Beloborodov 2017, 2019; Metzger et al.
2019), which can explain most observational properties of FRB
121102. For example, the persistent radio nebula associated with FRB
121102 can be produced by ion ejecta from the magnetar flares
(Beloborodov 2017). Meanwhile, both energy and particle content of
the nebula are consistent with this scenario, calibrated by
observations of ejecta from SGR 1806-20. Margalit \& Metzger (2018)
proposed that electron-ion nebula may explain the rotation measure
observed in FRB 121102.

\begin{figure*}
    \centering
    \subfigure{
        \label{fig:subfig:d}
        \hspace{7mm}\includegraphics[width=3.3in]{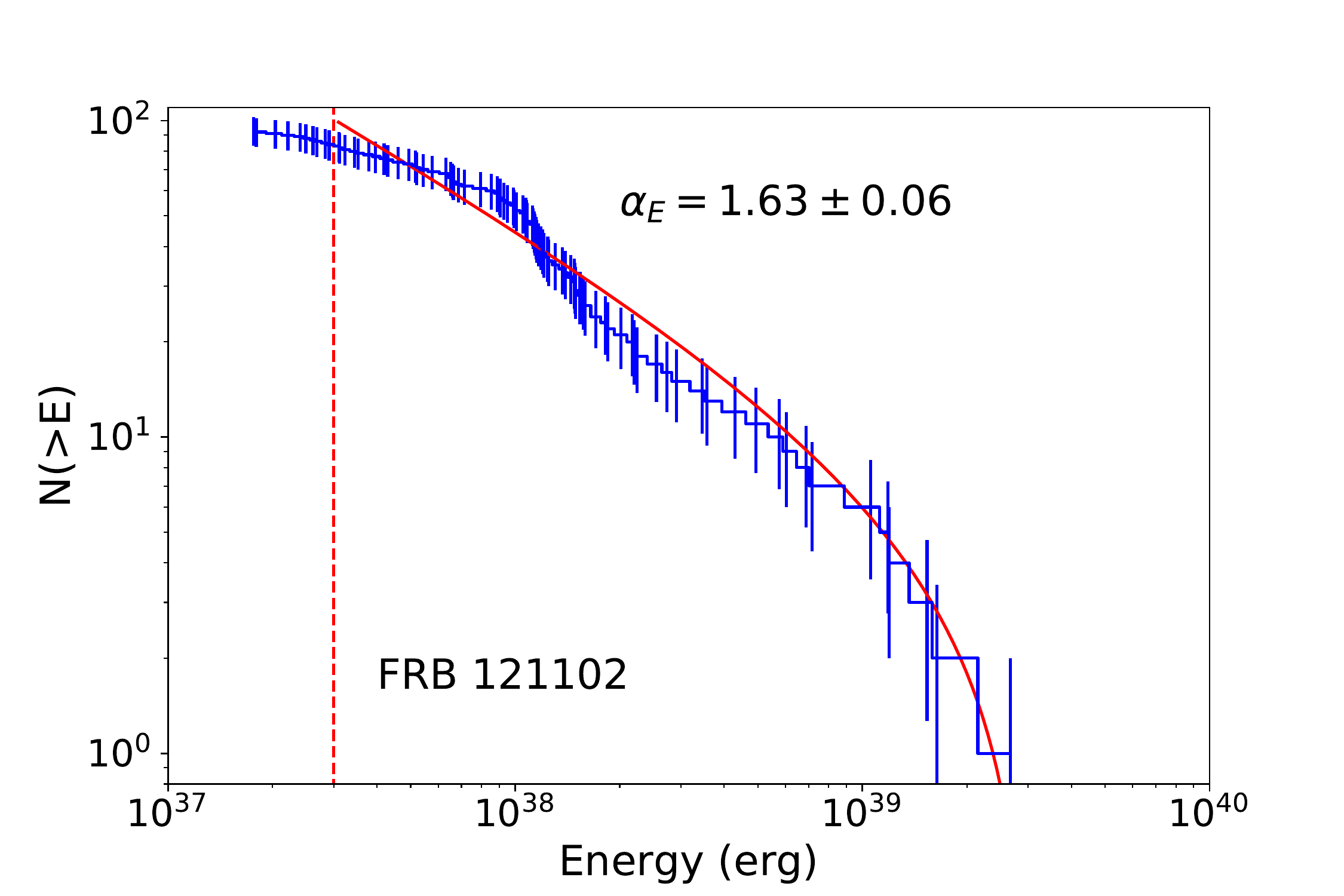}}
    \subfigure{
        \label{fig:subfig:e}
        \hspace{-8mm}\includegraphics[width=3.3in]{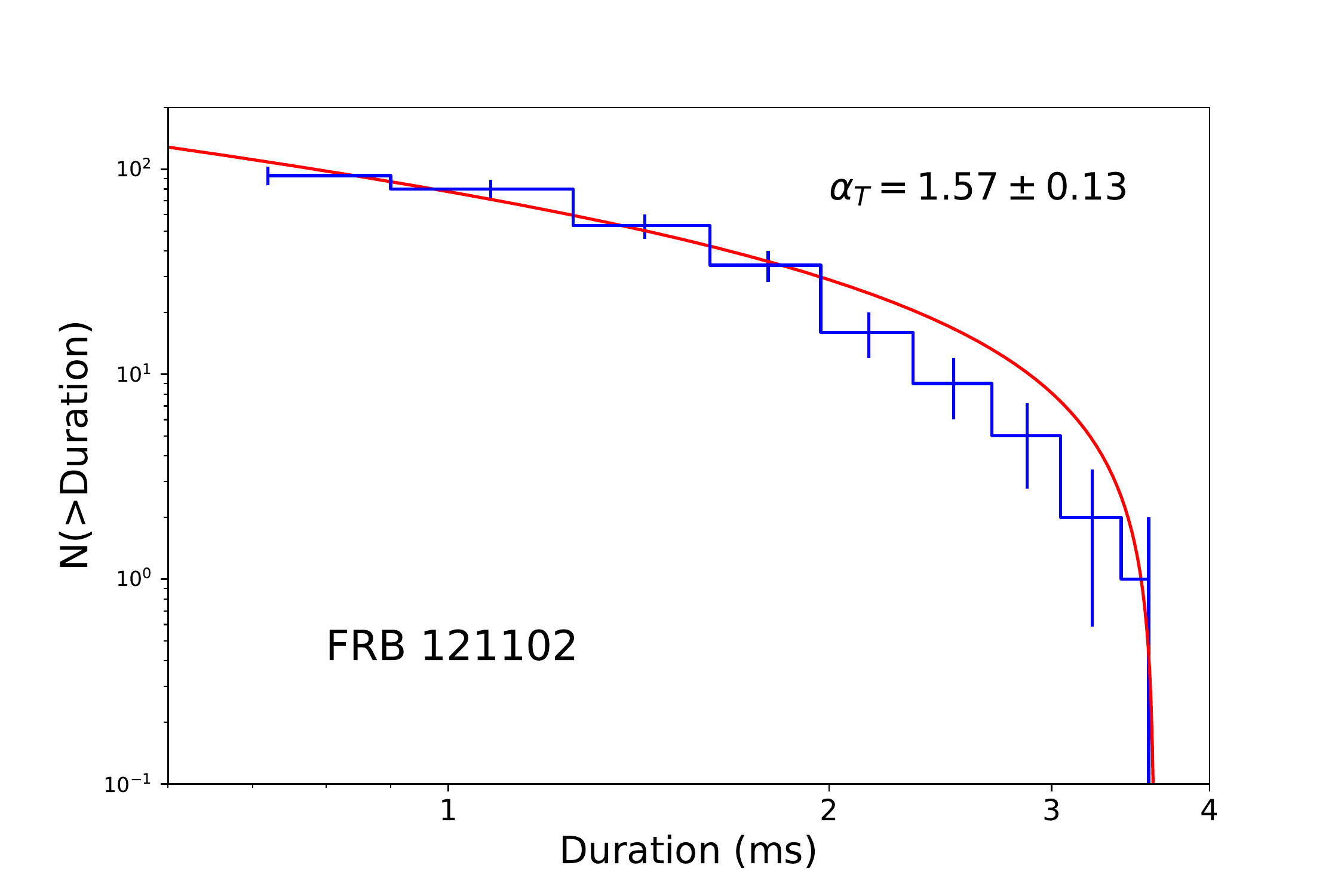}}
    \subfigure{
        \label{fig:subfig:f}
        \hspace{6mm}\includegraphics[width=3.3in]{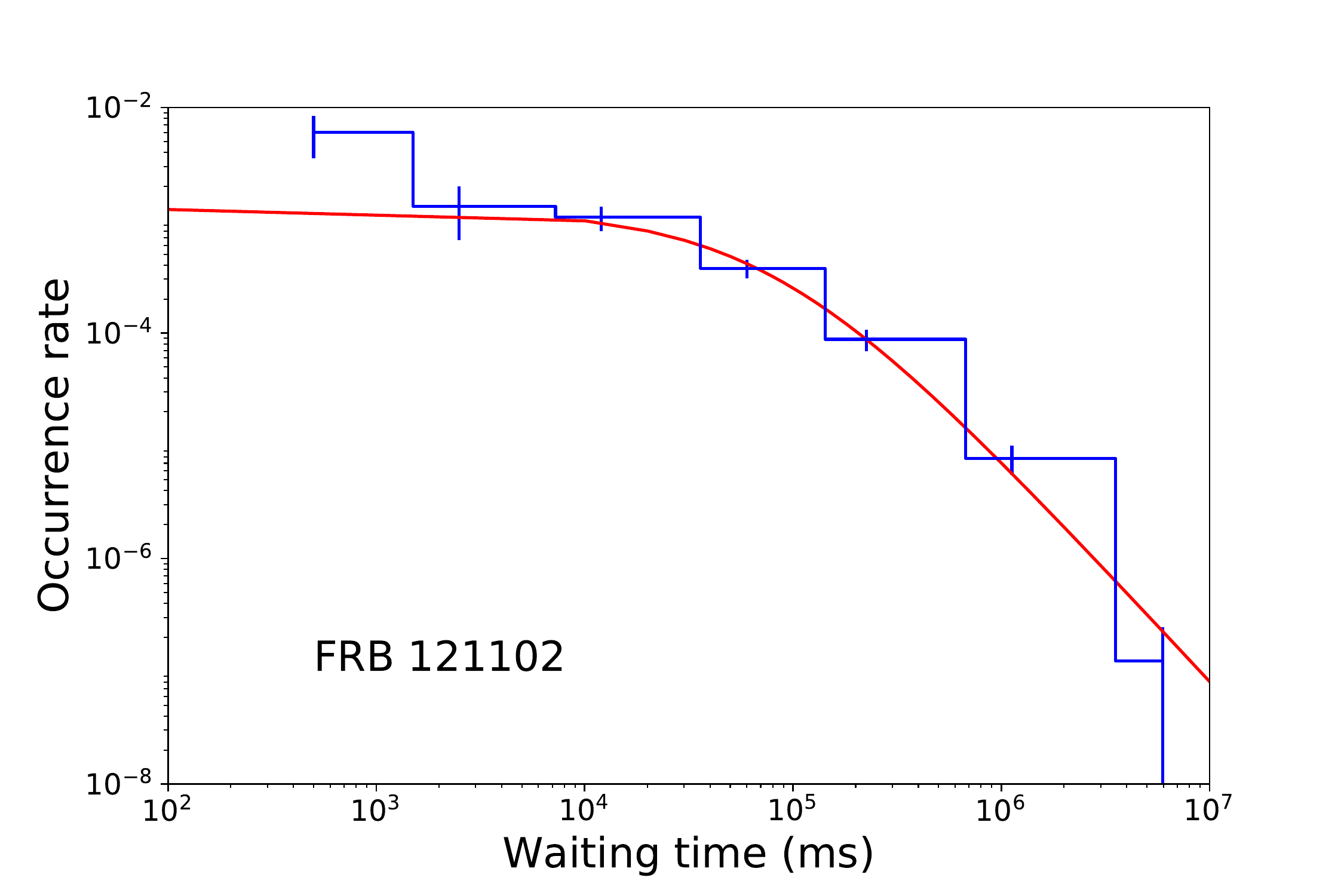}}
    \centering
    \caption{Frequency distributions of energy, duration and waiting time for FRB 121102. The best-fit results (red lines) are also shown.}
    \label{fig:5}
\end{figure*}

We use the largest sample of FRB 121102, which is observed by GBT at
4-8 GHz \citep{2018ApJ...866..149Z}. This sample contains 21 pulses
reported in \cite{2018ApJ...863....2G} and 72 pulses identified by
machine learning. These pulses were observed within a 6 hours
observation. They share the same observation conditions and were
observed by the same telescope. Therefore, we can put them together
to analysis and ignore complex selection effects. Power-law
distributions of energy $\alpha_E=1.63\pm0.06$ and distributions
$\alpha_T=1.57\pm0.13$ for these 93 FRB 121102 bursts are shown in
Figure \ref{fig:5}. Gourdji et al. (2019) discovered a
low-energy sample with 41 bursts for FRB 121102 and found
$\alpha_E\sim 1.7$ if all bursts are included (see their Figure 5).
However, if the low-energy bursts are discarded, a steeper
$\alpha_E\sim 2.8$ is found. Wang \& Zhang (2019) also found that
six samples of FRB 121102 bursts observed by different telescopes at
different frequencies show a universal energy distribution with
$\alpha_E$ around 1.7. Meanwhile, similar power-law index of energy
distribution for non-repeating FRBs is also found (Lu \& Piro 2019;
Zhang \& Wang 2019). The waiting time distribution of FRB
121102 also can be described by a non-stationary Poisson process
with mean burst rates $\lambda_{0}=1.23^{+0.80}_{-0.38} \times
10^{-5} ~\rm ms^{-1}$. Zhang et al. (2018) found that the rate of
detection is not stationary and the distribution of waiting time
cannot be well fitted using Poissonian distribution for the same
sample. For a small sample of waiting times of FRB 121102, Oppermann
et al. (2018) modeled the distribution of waiting times as Weibull
distribution, which can describe non-Poissonian distributions with
clustering. It must be noted that the non-stationary Poissonian
distribution used in this paper is similar to the Weibull
distribution. Because the rate of bursts in a non-stationary Poisson
process also varies with time (Wheatland et al. 1998). The mean
burst rate is 1.23$^{+0.80}_{-0.38}\times 10^{-2}$s$^{-1}$. Using
the same data, Zhang et al. (2018) found the rate is 0.05s$^{-1}$
for Poissonian distribution. Using a sparse waiting time sample,
Oppermann et al. (2018) derived a mean repetition rate of
$5.7^{+3.0}_{-2.0}$ day$^{-1}$. The large discrepancy between the
two rates is that the waiting times used in this paper are derived
from 93 bursts in 5 hr observation, comparing to 17 bursts in about
74 hr observation in Oppermann et al. (2018).

We also show the fitting results of energy, duration and waiting
time for FRB 121102 in Table 2. Therefore, similar distributions
between FRB 121102 and magnetar bursts are found. Meanwhile, Metzger
et al. (2019) discussed that FRBs could arise from synchrotron maser
emission at ultra-relativistic magnetized shocks, such as produced
by flare ejecta from young magnetars. This model can explain the
observational properties of FRBs, including burst duration, high
intrinsic linear polarization, an spectral energy distribution with
complex frequency structure, the downward evolution of frequency
structure in FRB 121102 and 180814.J0422+73, and time-varying
dispersion measure. More interestingly, their model has a testable
prediction. According to their analysis (equation (40) of Metzger et
al. (2019)), the luminosity of FRBs is
\begin{equation}
L_{\rm FRB} \approx 3\times 10^{42}f_{-3}E_{b,43}t_{-3}^{-1} \rm
erg~s^{-1},
\end{equation}
where $f$ is the radiative efficiency, and $E_b$ is the energy of
magnetar burst. From this equation, we can see that the luminosity
of FRB is linearly proportional to the energy of magnetar burst. Our
results support that the central engine of FRB 121102 is a magnetar.
More recently, a similar model basing on magnetar has been proposed
(Beloborodov 2019), which can also explain most features of FRBs.

The millisecond duration of FRBs requires a compact object origin,
i.e., neutron stars. From the energy budget, the repeating FRB
121102 must be powered by the magnetic energy (Lyutikov et al.
2017), which is similar to but more extreme than giant flares
produced by Galactic magnetars. The internal magnetic energy of
neutron star is
\begin{equation}
E_{\rm B}\approx 4\pi R^3_{\rm NS}/3 \times B^2/8\pi\approx 3\times
10^{45} \rm erg~ B^2_{14}, \label{eq:Eb}
\end{equation}
where $B$ is the magnetic field and $R_{\rm NS} = 12$ km is
the radius of the neutron star. The maximum number of bursts
produced by a given magnetar is
\begin{eqnarray}
N_{\rm FRB} &=& \frac{E_{\rm B}}{E_{\rm FRB}} \nonumber \\
&\approx& 0.03f_{b}^{-1}\left(\frac{f_{\rm
r}}{10^{-8}}\right)\left(\frac{B}{10^{16}\,{\rm
G}}\right)^{2}\left(\frac{E_{\rm FRB}}{10^{39}\rm erg}\right)^{-1},
\label{eq:Nburst}
\end{eqnarray}
where $f_r$ is the fraction of the flare energy placed into FRB
emission, $f_b$ is the beaming factor, and $E_{\rm FRB}=10^{39}$ erg is
the typical energy of FRB 121102 (Wang \& Zhang 2019). The FRB
efficiency is fixed to a value of $f_{\rm r} \approx 10^{-8}$
(Lyubarsky 2014). From observation, FRB 121102 has been active at
least 7 years with more than 200 bursts. From equation
(\ref{eq:Nburst}), the neutron star magnetic field has to be larger
than a threshold value:
\begin{equation}
B\geq 8\times 10^{15} \rm G ~f_{b}^{1/2}\left(\frac{f_{\rm
r}}{10^{-8}}\right)^{-1/2}\left(\frac{E_{\rm FRB}}{10^{39}\rm
erg}\right)^{1/2}.
\end{equation}
Therefore, for a large range of $f_{b}\geq 1.5\times 10^{-4}$, the
central neutron star must be a magnetar ($B\geq 10^{14}$G). Although
the value of $f_b$ is uncertain, it cannot be too small, which
causes that the beaming-correct FRB rate (proportional to
$f_b^{-1}$) would greatly exceed the rate of known objects (Zhang
2016), such as core-collapse supernova (Dahlen et al. 2004), binary
neutron star merger (Abbott et al. 2017), long gamma-ray bursts (Yu
et al. 2015) and short gamma-ray bursts (Zhang \& Wang 2018).

It is important to search radio bursts associated with high-energy
bursts of magnetars. Interestingly, XTE J1810-197 is the first ever
magnetar emitting transient radio burst. More recently, the second
radio outburst that has been observed from this magnetar (Maan et
al. 2019). The bursts show a characteristic intrinsic width of the
order of 0.5-0.7 ms. It is also found that the bursts exhibit
possible similar spectral structures to that of FRB 121102 (Maan et
al. 2019). The magnetar J1810-197 is only the third object after
repeating FRBs and the Crab pulsar which is found to exhibit
frequency structures, may provide a link between the underlying
emission mechanisms for magnetars and repeating FRBs. However,
higher time-resolution of observation is required.

\section{Conclusions}
\label{sec:discuss} In this paper, we find power-law distributions
of total counts and durations for bursts from three magnetars.
Power-law energy distributions have also been found for solar flares
with $\alpha_E$ = $1.53-1.73$ (Crosby et al. 1993; Lu et al. 1993;
Aschwanden 2011), for flares from gamma-ray bursts with
$\alpha_E=1.06$ (Wang \& Dai 2013; Yi et al. 2016, 2017), for flares
from black holes with $\alpha_E=1.6-2.1$ (Wang et al. 2015; Yan et
al. 2018). Power-law duration distributions also found in above
systems. This is a typical behavior for SOC systems. The concept of
SOC (Bak et al. 1987) states that subsystems self-organize due to
some driving force to a critical state at which a slight
perturbation can cause a chain reaction of any size within the
system. Magnetar power-law energy and duration distributions, along
with a non-stationary waiting time distribution, support that
systems responsible for magnetar bursts are in a SOC state. For
magnetars, the critical systems are neutron star crusts strained by
evolving magnetic stresses (Thompson \& Duncan 1995). We also found
similar energy, duration and waiting time distributions between
magnetars and FRB 121102, together with some theoretical models of
FRBs basing on magnetar, which indicate that the central engine of
FRB 121102 is a magnetar. In future, much more repeating FRBs will
be discovered. The connection between repeating FRBs and magnetars
can be tested.

\section*{Acknowledgements}
We greatly acknowledge two anonymous referees for the valuable
comments, which have significantly improved our work. This work is
supported by the National Natural Science Foundation of China (grant
U1831207).

\begin{thebibliography}{}
\bibitem[Abbott et al. (2017)]{Abbott2017}Abbott, B. P., Abbott, R., Abbott, T. D., et al. 2017, ApJL, 848, L13
\bibitem[Aschwanden(2011)]{2011soca.book.....A} Aschwanden, M.~J.\ 2011, Self-Organized Criticality in Astrophysics
\bibitem[Aschwanden(2015)]{2015ApJ...814...19A} Aschwanden, M.~J.\ 2015, \apj, 814, 19
\bibitem[Aschwanden \& McTiernan (2010)]{2010ApJ...717...683A} Aschwanden, M.~J. \& McTiernan, J. M., 2010, \apj, 717, 683
\bibitem[Astropy Collaboration et al.(2013)]{2013A&A...558A..33A} Astropy Collaboration, Robitaille, T.~P., Tollerud, E.~J., et al.\ 2013, \aap, 558, A33
\bibitem[Bak et al.(1987)]{1987PhRvL..59..381B} Bak, P., Tang, C., \& Wiesenfeld, K.\ 1987, \prl, 59, 381
\bibitem[Bak \& Tang(1989)]{1989JGR....9415635B} Bak, P., \& Tang, C.\ 1989, Journal of Geophysical Research, 94, 15,635
\bibitem[Beloborodov (2017)]{Beloborodov2017} Beloborodov, A. M. 2017, ApJL, 843, L26
\bibitem[Beloborodov (2019)]{Beloborodov2019} Beloborodov, A. M. 2019, arXiv:1908.07743
\bibitem[Bannister et al.(2019)]{Bannister2019} Bannister, K. W., Deller, A. T., Phillips, C., et al. 2019, Science, 365, 565
\bibitem[Bour \& Davy(1997)]{1997WRR....33.1567B} Bour, O., \& Davy, P.\ 1997, Water Resources Research, 33, 1567
\bibitem[Cao et al. (2017)]{Cao2017}Cao, X.-F., Yu, Y.-W., \& Dai, Z.-G. 2017, ApJL, 839, L20
\bibitem[Chatterjee et al.(2017)]{Chatterjee2017} Chatterjee, S., Law, C. J., Wharton, R. S., et al. 2017, Nature, 541, 58
\bibitem[Cheng et al.(1996)]{1996Natur.382..518C} Cheng, B., Epstein, R.~I., Guyer, R.~A., et al.\ 1996, Nature, 382, 518
\bibitem[CHIME/FRB Collaboration (2019a)]{Amiri2019} CHIME/FRB Collaboration, Amiri, M., Bandura, K., et al.\ 2019, Nature, 566, 235
\bibitem[CHIME/FRB Collaboration (2019b)]{Andersen2019}CHIME/FRB Collaboration, Andersen, B. C., et al.\ 2019, arXiv:1908.03507
\bibitem[Cordes \& Chatterjee (2019)]{Cordes2019}Cordes, J. M., \& Chatterjee, S. 2019, ARA\&A, 57, 417
\bibitem[Crosby et al. (1993)]{Crosby1993} Crosby, N. B., Aschwanden, M., \& Dennis, B. 1993, Sol. Phys., 143, 275
\bibitem[Dahlen et al. (2014)]{Dahlen2004} Dahlen, T., Strolger, L. G., Riess, A., et al. 2004, ApJ, 613, 189
\bibitem[Gajjar et al.(2018)]{2018ApJ...863....2G} Gajjar, V., Siemion, A.~P.~V., Price, D.~C., et al.\ 2018, \apj, 863, 2
\bibitem[G{\"o}{\v{g}}{\"u}{\c{s}} et al.(1999)]{Gogus1999}G{\"o}{\v{g}}{\"u}{\c{s}}, E., Woods, P.~M., Kouveliotou, C., et al.\ 2000, \apj, 526, L93
\bibitem[G{\"o}{\v{g}}{\"u}{\c{s}} et al.(2000)]{2000ApJ...532L.121G} G{\"o}{\v{g}}{\"u}{\c{s}}, E., Woods, P.~M., Kouveliotou, C., et al.\ 2000, \apj, 532, L121
\bibitem[G{\"o}{\v{g}}{\"u}{\c{s}} et al.(2001)]{2001ApJ...558..228G} G{\"o}{\v{g}}{\"u}{\c{s}}, E., Kouveliotou, C., Woods, P.~M., et al.\ 2001, \apj, 558, 228
\bibitem[Gourdji et al.(2019)]{Gourdji19} Gourdji, K., Michilli, D., Spitler, L.~G., et al.\ 2019, ApJL, 877, L19
\bibitem[Harko et al. (2015)]{Harko2015}Harko, T., Mocanu, G. \& Stroia, N., 2015, Astrophysics and Space Science, 357, 84
\bibitem[Hurley et al. (1994)]{Hurley1994} Hurley, K. J., et al., 1994, A\&A, 288, L49
\bibitem[Israel et al.(2008)]{2008ApJ...685.1114I} Israel, G.~L., Romano, P., Mangano, V., et al.\ 2008, \apj, 685, 1114
\bibitem[Katz(1996)]{1996ApJ...463..305K} Katz, J.~I.\ 1996, \apj, 463, 305
\bibitem[Katz(2016)]{2016ApJ...826..226K} Katz, J.~I.\ 2016, \apj, 826, 226
\bibitem[K{\i}rm{\i}z{\i}bayrak et al. (2017)]{kirm2017} K{\i}rm{\i}z{\i}bayrak, D., et al. 2017, ApJS, 232, 17
\bibitem[Kouveliotou(1995)]{1995Ap&SS.231...49K} Kouveliotou, C.\ 1995, \apss, 231, 49
\bibitem[Kulkarni et al.(2014)]{2014ApJ...797...70K} Kulkarni, S.~R., Ofek, E.~O., Neill, J.~D., et al.\ 2014, \apj, 797, 70
\bibitem[Li et al.(2015)]{Li2015} Li, Y. P., et al.\ 2015, \apj, 810, 19
\bibitem[Lin et al.(2011)]{2011ApJ...740L..16L} Lin, L., Kouveliotou, C., G{\"o}{\v{g}}{\"u}{\c{s}}, E., et al.\ 2011, \apj, 740, L16
\bibitem[Lin et al.(2013)]{Lin2013} Lin, L., Gogus, E., Kaneko, Y., et al. 2013, ApJ, 778, 105
\bibitem[Lorimer et al.(2007)]{2007Sci...318..777L} Lorimer, D.~R., Bailes, M., McLaughlin, M.~A., et al.\ 2007, Science, 318, 777
\bibitem[Lu \& Hamilton(1991)]{1991ApJ...380L..89L} Lu, E.~T., \& Hamilton, R.~J.\ 1991, \apj, 380, L89
\bibitem[Lu et al. (1993)]{Lu1993} Lu, E., Hamilton, R., McTiernan, J., \& Bromund, K. 1993, ApJ, 412, 841
\bibitem[Lu \& Kumar(2018)]{Lu2018MNRAS.477.2470L} Lu, W., \& Kumar, P.\ 2018, \mnras, 477, 2470
\bibitem[Lu \& Piro (2019)]{Lu2019}Lu, W., \& Piro, A. L. 2019, ApJ, 883, 40
\bibitem[Lyubarsky (2014)]{Lyu2014} Lyubarsky, Y. 2014, MNRAS, 442, L9
\bibitem[Lyutikov (2017)]{Lyutikov2017} Lyutikov, M. 2017, ApJL, 838, L13
\bibitem[Maan et al. (2019)]{Maan2019} Maan, Y. et al., 2019, arXiv: 1908.04304
\bibitem[Marcote et al. (2107)]{Marcote2017} Marcote, B., Paragi, Z., Hessels, J. W. T., et al. 2017, ApJL, 834, L8
\bibitem[Margalit \& Metzger (2018)]{Margalit2018} Margalit, B., \& Metzger, B. D. 2018, ApJ, 868, L4
\bibitem[Mereghetti(2008)]{2008A&ARv..15..225M} Mereghetti, S.\ 2008, Astronomy and Astrophysics Review, 15, 225
\bibitem[Metzger et al. (2017)]{Metzger2017}Metzger, B. D., Berger, E., \& Margalit, B. 2017, ApJ, 841, 14
\bibitem[Metzger et al. (2019)]{Metzger2019} Metzger, B. D., Margalit, B., \& Sironi, L., 2019, MNRAS, 485, 4091
\bibitem[Michilli et al. (2018)]{Michilli2018}Michilli, D., Seymour, A., Hessels, J. W. T., et al. 2018, Nature, 533, 132
\bibitem[Nakagawa et al.(2007)]{2007PASJ...59..653N} Nakagawa, Y.~E., Yoshida, A., Hurley, K., et al.\ 2007, PASJ, 59, 653
\bibitem[Newman(2005)]{2005ConPh..46..323N} Newman, M.~E.~J.\ 2005, Contemporary Physics, 46, 323
\bibitem[Olami et al.(1992)]{1992PhRvL..68.1244O} Olami, Z., Feder, H.~J.~S., \& Christensen, K.\ 1992, \prl, 68, 1244
\bibitem[Olausen \& Kaspi(2014)]{2014ApJS..212....6O} Olausen, S.~A., \& Kaspi, V.~M.\ 2014, ApJS, 212, 6
\bibitem[Opperman et al. (2018)]{Opperman18}Opperman, N., Yu, H. \& Pen, U.-L. 2018, MNRAS, 475, 5109
\bibitem[Petroff et al. (2016)]{Petroff16}Petroff, E., Barr, E. D., Jameson, A., et al. 2016, PASA, 33, e045
\bibitem[Petroff et al. (2019)]{Petroff19}Petroff, E., Hessels, J.W.T. \& Lorimer, D.R., 2019, Astron Astrophys Rev, 27, 4
\bibitem[Pinto et al.(2012)]{2012CNSNS..17.3558P} Pinto, C.~M.~A., Mendes Lopes, A., \& Machado, J.~A.~T.\ 2012, Communications in Nonlinear Science and Numerical Simulations, 17, 3558
\bibitem[Platts et al.(2019)]{Platts2018}Platts, E., Weltman, A., Walters, A., et al. 2019, Physics Reports, 821, 1
\bibitem[Popov (2013)]{Popov2013} Popov, S. B., \& Postnov, K. A. 2013, arXiv:1307.4924
\bibitem[Prieskorn \& Kaaret(2012)]{2012ApJ...755....1P} Prieskorn, Z., \& Kaaret, P.\ 2012, \apj, 755, 1
\bibitem[Ravi et al.(2019)]{Ravi2019} Ravi, V., Catha, M., D'Addario, L., et al. 2019, arXiv:1907.01542
\bibitem[Rosner \& Vaiana(1978)]{1978ApJ...222.1104R} Rosner, R., \& Vaiana, G.~S.\ 1978, \apj, 222, 1104
\bibitem[Salvatier et al.(2016)]{2016ascl.soft10016S} Salvatier, J., Wiecki{\^a}, T.~V., \& Fonnesbeck, C.\ 2016, PyMC3: Python probabilistic programming framework, ascl:1610.016
\bibitem[Scargle et al.(2013)]{Scargle2013} Scargle, J. D., Norris, J. P., Jackson, B., et al. 2013, ApJ, 764, 167
\bibitem[Schwarz et al.(2011)]{2011ApJS..197...31S} Schwarz, G.~J., Ness, J.-U., Osborne, J.~P., et al.\ 2011, \apjs, 197, 31
\bibitem[Spitler et al.(2016)]{2016Natur.531..202S} Spitler, L.~G., Scholz, P., Hessels, J.~W.~T., et al.\ 2016, Nature, 531, 202.
\bibitem[Suvorov \& Kokkotas(2019)]{Suvorov2019}Suvorov, A. G. \& Kokkotas, K. D., 2019, MNRAS, 488, 5887
\bibitem[Thompson \& Duncan(1995)]{1995MNRAS.275..255T} Thompson, C., \& Duncan, R.~C.\ 1995, \mnras, 275, 255
\bibitem[Thornton et al.(2013)]{2013Sci...341...53T} Thornton, D., Stappers, B., Bailes, M., et al.\ 2013, Science, 341, 53.
\bibitem[van der Horst et al.(2012)]{2012ApJ...749..122V} van der Horst, A.~J., Kouveliotou, C., Gorgone, N.~M., et al.\ 2012, \apj, 749, 122
\bibitem[Vogt et al.(2014)]{2014ApJ...793..127V} Vogt, F.~P.~A., Dopita, M.~A., Kewley, L.~J., et al.\ 2014, \apj, 793, 127
\bibitem[Wang \& Dai(2013)]{Wang2013NatPh...9..465W} Wang, F.~Y., \& Dai, Z.~G.\ 2013, Nature Physics, 9, 465.
\bibitem[Wang et al.(2015)]{Wang2015ApJS..216....8W} Wang, F.~Y., Dai, Z.~G., Yi, S.~X., et al.\ 2015, ApJS, 216, 8.
\bibitem[Wang \& Yu(2017)]{2017JCAP...03..023W} Wang, F.~Y., \& Yu, H.\ 2017, JCAP, 03, 023
\bibitem[Wang \& Zhang(2019)]{Wang2019} Wang, F.~Y., \& Zhang, G. Q. 2019, ApJ, 882, 108
\bibitem[Wang et al.(2017)]{WangJ2017} Wang, J. S., Wang, F.~Y. \& Dai, Z.~G., 2017, MNRAS, 471, 2517
\bibitem[Wheatland et al.(1998)]{1998ApJ...509..448W} Wheatland, M.~S., Sturrock, P.~A., \& McTiernan, J.~M.\ 1998, \apj, 509, 448
\bibitem[Wheatland \& Litvinenko(2002)]{2002SoPh..211..255W} Wheatland, M.~S., \& Litvinenko, Y.~E.\ 2002, \solphys, 211, 255
\bibitem[Willis \& Yule(1922)]{1922Natur.109..177W} Willis, J.~C., \& Yule, G.~U.\ 1922, Nature, 109, 177
\bibitem[Woods \& Thompson(2006)]{2006csxs.book..547W} Woods, P.~M., \& Thompson, C.\ 2006, Compact Stellar X-ray Sources, 547
\bibitem[Woods et al. (1999)]{Woods1999} Woods, P., et al. 1999, ApJ, 524, L55
\bibitem[Yan et al. (2018)]{Yan2018} Yan, D. H., et al. 2018, ApJ, 864, 164
\bibitem[Yang \& Zhang (2018)]{Yang2018} Yang, Y. P. \& Zhang, B. 2018, ApJ, 868, 31
\bibitem[Yi et al. (2016)]{Yi2016} Yi, S. X., et al. 2016, ApJS, 224, 20
\bibitem[Yi et al. (2017)]{Yi2017} Yi, S. X., et al., 2017, ApJ, 844, 79
\bibitem[Yu et al. (2015)]{Yu2015} Yu, H., Wang, F. Y., Dai, Z. G., et al. 2015, ApJS, 218, 13
\bibitem[Zhang(2016)]{Zhang2016}Zhang, B., 2016, ApJL, 822, L14
\bibitem[Zhang \& Wang (2018)]{Zhang2018}Zhang, G. Q., \& Wang, F. Y. 2018, ApJ, 852, 1
\bibitem[Zhang et al.(2018)]{2018ApJ...866..149Z} Zhang, Y.~G., Gajjar, V., Foster, G., et al.\ 2018, \apj, 866, 149
\end{thebibliography}
\end{document}